\documentstyle[]{article}
\begin{document}

\small \hoffset = -1truecm \voffset = -2truecm
\title{\bf On the tachyon inflation}
\author{
    Xin-zhou Li\footnote {e-mail address: kychz@shtu.edu.cn}\hspace{0.6cm}  Dao-jun Liu \hspace{0.6cm} Jian-gang Hao \\\footnotesize \it
    Shanghai United Center for Astrophysics, Shanghai Normal University, Shanghai 200234
    ,China\\\footnotesize \it
Institute for Theoretical Physics, East China University of
Science and Technology, Shanghai 200237, China}
\date{}
\maketitle
\begin{abstract}
  Although the formulas of tachyon inflation correspond to those
of the inflation driven by the ordinary scalar field, there is
obvious difference between them, which can not be neglected. We
calculate the scalar and tensor perturbation of the string theory
inspired tachyon inflation, which has been widely studied
recently. We also show, through the Hamilton-Jacobi approach, that
the the rolling tachyon can essentially produce enough inflation.
An exact solution with the inverse squared potential of tachyon
field has been proposed and its power spectra has been analyzed.
\end{abstract}
\hspace{0.8cm} PACS numbers: 98.80.Cq, 98.80.Hw, 11.25.Sq

\section{Introduction}

The idea of inflation is legitimately regarded as an great
advancement of modern cosmology: it solves the horizon, flatness
and monopole problem, and it provides a mechanism for the
generation of density perturbations needed to seed the formation
of structures in the universe \cite{Kolb}. In standard
inflationary models \cite{Liddle}, the physics lies in the
inflation potential. The underlying dynamics is simply that of a
single scalar field rolling in its potential. This scenario is
generically referred to as chaotic inflation in reference to its
choice of initial conditions. This picture is widely favored
because of its simplicity and has received by far the most
attention to date. Some potentials that give the correct
inflationary properties have been proposed \cite{Turner} in the
past two decades. However, most of these investigations were not a
stickler for the fundamental physics, such as the standard model
or grand unified theories in particle physics. The above-mentioned
conventional physics does not yield an inflation potential that
agrees with observations, such as enough number of e-folds, the
amplitude of the density perturbation and etc.

Recently, pioneered by Sen \cite{Sen1}, the study of non-BPS
objects such as non-BPS branes, brane-antibrane configurations or
space-like branes \cite{Strominger} has attracted physical
interests in string theory. Sen showed that classical decay of
unstable D-brane in string theories produces pressureless gas with
non-zero energy density \cite{Sen2}. Gibbons took into account the
gravitational coupling by adding an Einstein-Hilbert term to the
effective action of the tachyon on a brane, and initiated a study
of "tachyon cosmology" \cite{Gibbons}. The basic idea is that the
usual open string vacuum is unstable but there exists a stable
vacuum with zero energy density. There is evidence that this state
is associated with the condensation of electric flux tubes of
closed string \cite{Sen2}. These flux tubes are successfully
described by using an effective Born-Infeld action \cite{Sen3}.
The string theory motivated tachyon inflation has been discussed
in Ref.\cite{Fairbairn, Kofman}. These investigation are based on
the slow-roll approximation so that they are rather incomplete.
Slow-roll is not the only possibility for successfully realistic
models of inflation, and solutions outside the slow-roll
approximation have been found in particular situations
\cite{Stewart}. The elegant review of Hamilton-Jacobi formalism in
inflationary cosmology with ordinary scalar field can be found in
Ref.\cite{Lidsey}.

In this paper, we discuss the generic properties of tachyon
inflation with the Born-Infeld action. Although the formulas of
tachyon inflation are correspond to those of the inflation driven
by the ordinary scalar field, there is obvious difference between
them. We also show an exact solution and a string motivated model.



\section{Rolling tachyon dynamics}
\subsection{The general formalism}

We consider spatially flat FRW line element given by:

\begin{eqnarray}\label{bgMetric}
ds^2=& &dt^2-a^2(t)(dx^2+dy^2+dz^2)\nonumber\\
    =& &a^2(\tau)[d\tau^2-(dx^2+dy^2+dz^2)]
\end{eqnarray}

\noindent where $\tau$ is the conformal time, with $dt=ad\tau$. As
shown by Sen\cite{Sen2}, a rolling tachyon condensate in either
bosonic or supersymmetric string theory can be described by a
fluid which, in the homogeneous limit, has energy density and
pressure as follows

\begin{equation}
\rho=\frac{V(T)}{\sqrt{1-\dot{T}^2}},\hspace{0.5cm}
p=-V(T)\sqrt{1-\dot{T}^2}
\end{equation}

\noindent where $T$ and $V(T)$ are the tachyon field and
potential, and an overdot denotes a derivative with respect to the
coordinate time $t$. When taking the gravitational field into
account, the effective Lagrangian density in the Born-Infeld-type
is \cite{Gibbons}

\begin{equation}\label{lagrangian}
L=\sqrt{-g}(\frac{R}{2\kappa}-V(T)\sqrt{1-g^{\mu\nu}\partial_\mu
T\partial_\nu T})
\end{equation}

\noindent where $\kappa=8\pi G=Mp^{-2}$. For a spatially
homogenous tachyon field $T$, we have the equation of motion

\begin{equation}\label{ddT}
\ddot{T}+3H\dot{T}(1-\dot{T}^2)+\frac{V'}{V}(1-\dot{T}^2)=0
\end{equation}

\noindent which is equivalent to the entropy conservation
equation. Here, the Hubble parameter $H$ is defined as $H\equiv
(\frac{\dot{a}}{a})$, and $V'=dV/dT$. If the stress-energy of the
universe is dominated by the tachyon field $T$, the Einstein field
equations for the evolution of the background metric,
$G_{\mu\nu}=\kappa T_{\mu\nu}$, can be written as

\begin{equation}\label{H2}
H^2=(\frac{\dot{a}}{a})^2=\frac{\kappa}{3}\frac{V(T)}{\sqrt{1-\dot{T}^2}}
\end{equation}

\noindent and

\begin{equation}\label{dda}
(\frac{\ddot{a}}{a})=H^2+\dot{H}=\frac{\kappa}{3}\frac{V(T)}{\sqrt{1-\dot{T}^2}}(1-\frac{3}{2}\dot{T}^2)
\end{equation}

\noindent Eqs.(\ref{ddT})-(\ref{dda}) form a coupled set of
evolution equations of the universe. The fundamental quantities to
be calculated are $T(t)$ and $a(t)$, and the potential $V(T)$ is
given when the model is specified. The period of accelerated
expansion corresponds to $\dot{T}^2<\frac{3}{2}$ and decelerate
otherwise. In the limit case $\dot{T}=0$, there is no difference
in meaning of the expansion of universe between tachyon inflation
and ordinary inflation driven by inflaton. However, the case of
$\dot{T}\neq 0$ forms a sharp contrast. Although the formulas of
tachyon inflation are correspond to those of the inflation driven
by ordinary scalar field, there is obvious difference between them
which can not be neglected. From Eqs.(\ref{ddT})-(\ref{dda}), we
have two first-order equations

\begin{equation}
\dot{T}=-\frac{2}{3}\frac{H'(T)}{H^2(T)}
\end{equation}

\begin{equation}\label{HJE:8}
 [H'(T)]^2-\frac{9}{4}H^4(T)=-\frac{\kappa^2}{4}V^2(T)
\end{equation}

\noindent These equations are wholly equivalent to the
second-order equation of motion(\ref{ddT}).

Analogous to the inflation driven by ordinary scalar field
\cite{Lidsey}, we define the "slow-roll" parameters as follows

\begin{equation}\label{epsilon:9}
\epsilon(T)\equiv \frac{2}{3}\bigg(\frac{H'(T)}{H^2(T)}\bigg)^2
\end{equation}

\begin{equation}
\eta(T)\equiv\frac{1}{3}\bigg(\frac{H''(T)}{H^3(T)}\bigg)
\end{equation}

\noindent and

\begin{equation}\label{xi:11}
\xi(T)\equiv
\frac{2}{3}\bigg(\frac{H'(T)H'''(T)}{H^6(T)}\bigg)^{\frac{1}{2}}
\end{equation}

\noindent where the sign ambiguity is the result of the convention
that $\sqrt{\epsilon}$ is always taken to be positive. Clearly,
the definitions of the parameters
Eqs.(\ref{epsilon:9})-(\ref{xi:11}) are quite different from those
defined in ordinary inflation. This is very natural for the
Born-Infeld action is sharply different from that of the ordinary
scalar field. In term of $\epsilon$ parameter, Eq.(\ref{HJE:8})
can be reexpressed as

\begin{equation}\label{vh:12}
H^4(T)[1-\frac{1}{3}\epsilon (T)]=\frac{\kappa^2}{9}V^2(T)
\end{equation}

\noindent which is referred to as the Hamilton-Jacobi equation of
tachyon inflation. The number of e-folds of the inflation produced
when the tachyon field rolls from a particular value$T$ to the end
point $T_e$ is

\begin{equation}
N(T,T_e)\equiv \int^{t_e}_t
H(t)dt)=\int^{T_e}_{T}\frac{H}{\dot{T}}dT
\end{equation}

\noindent Therefore, we have

\begin{equation}
a(T)=a_e \exp[-N(T)]
\end{equation}

\noindent where $a_e$ is the value of the scale factor at the end
of inflation. To match the observed degree of flatness and
homogeneity in the universe, one requires many e-folds of
inflation, typically $N\approx 50$.

During inflation, the tachyon and graviton fields underwent
quantum fluctuations. The most important observational deductions
of the inflationary scenario is that inflation explains not only
the high degree of large-scale isotropy in the universe, but also
the underlying mechanism for the observed anisotropy. The quantum
fluctuations on small scales are quickly redshifted to scales much
larger than the horizon, where they are "congealed" as
perturbations in the background metric during inflation epoch.

It is not difficult to prove that the gravitational and curvature
perturbations in the tachyon inflation take similar form as those
in the conventional inflation, except the definitions of the
parameters. Therefore, we will carry out our analysis in analogy
with the discussions for ordinary inflation
\cite{Lidsey,Bardeen,Grishchuk,Mukhanov1,Mukhanov2}. The spectrum
of curvature perturbation $P_R(k)$ as function of wavenumber $k$
could be expressed as \cite{Mukhanov1}

\begin{equation}\label{PRk:17}
P^{\frac{1}{2}}_R(k)=\sqrt{\frac{k^3}{2\pi^2}}\mid
\frac{u_k}{z}\mid
\end{equation}

\noindent where mode function $u_k$ satisfies following equation
\cite{Grishchuk}

\begin{equation}\label{Eq:uk}
\frac{d^2u_k}{d\tau^2}+(k^2-\frac{1}{z}\frac{d^2z}{d\tau^2})u_k=0
\end{equation}

\noindent and the quantity $z$ is defined as

\begin{equation}\label{z}
z\equiv \frac{a\dot{T}}{H}
\end{equation}

\noindent and

\begin{equation}\label{hh}
\frac{1}{z}\frac{d^2z}{d\tau^2}=2a^2H^2(1+4\epsilon -3\eta
+9\epsilon^2-14\epsilon \eta+2\eta^2+\frac{1}{2}\xi^2)
\end{equation}

\noindent Clearly, the above expression Eq.(\ref{hh}) is different
from that of the ordinary scalar field because the coupling of
curvature perturbations to the stress-energy of tachyon field is
in a very different manner.

Similarly, the spectrum of gravitational wave $P_g(k)$ is defined
by

\begin{equation}\label{Pgk:21}
P^{\frac{1}{2}}_g(k)=\frac{4}{m_{pl}}\sqrt{\frac{k^3}{\pi
a^2}}\mid v_{k}\mid
\end{equation}

\noindent where mode function $v_k$ satisfies \cite{Grishchuk}

\begin{equation}\label{Eq:vk}
\frac{d^2v_k}{d\tau^2}+(k^2-\frac{a_{\tau \tau }}{a})v_k=0
\end{equation}

\noindent where

\begin{equation}\label{tt}
\frac{a_{\tau \tau }}{a}=2a^2H^2(1-\frac{\epsilon}{2})
\end{equation}

\noindent The above expression Eq.(\ref{tt}) is the same as that
in ordinary scalar field inflation, because the tensor
perturbations only describe the propagation of gravitational waves
and do not couple to the matter term.

\subsection{Exact solution for power-law inflation of tachyon}

In this subsection, we will analyze an exactly solvable model in
tachyon inflation. The potential in this model is the inverse
square potential

\begin{equation}
V(T)=\frac{2}{\kappa}(n^2-\frac{n}{3})^{\frac{1}{2}}(T-T_0)^{-2}
\end{equation}

\noindent which played an important role in the inflationary
cosmology of tachyon. The spectrum equations (\ref{Eq:uk}) and
(\ref{Eq:vk}) with the inverse square potential can be solved
exactly. This potential corresponds to the well known Power-law
inflation, in which the scale factor expands as $a(t)\propto t^n$.
In this case, the parameters take the particularly simple forms:

\begin{equation}\label{H:25}
H(T)=\frac{(\frac{2}{3}n)^{\frac{1}{2}}}{T-T_0}
\end{equation}

\noindent and

\begin{equation}\label{SlowRollParameters}
\epsilon= \frac{1}{n}, \hspace{0.5cm} \eta=
\frac{1}{n},\hspace{0.5cm} \xi=\frac{\sqrt{6}}{n}.
\end{equation}

According to Eq.(\ref{H:25}), the spectrum equations for curvature
perturbations and gravitational perturbations are reduced to the
following Bessel equations respectively,

\begin{equation}
[\frac{d^2}{d\tau^2}+k^2-\frac{\mu^2-\frac{1}{4}}{\tau^2}]u_k=0
\end{equation}

\begin{equation}
[\frac{d^2}{d\tau^2}+k^2-\frac{\nu^2-\frac{1}{4}}{\tau^2}]v_k=0
\end{equation}

\noindent where $\mu=\frac{3}{2}+\frac{2}{n-1}$ and
$\nu=\frac{3}{2}+\frac{1}{n-1}$.  It is worth noting that the mode
functions with $\mu\neq \nu$ in the tachyon inflation is quite
different from that in the power-law inflation driven by ordinary
scalar field, in which $\mu=\nu$. In the long wavelength limit,
$k/aH \rightarrow 0$, the asymptotic forms of the power spectrum
are

\begin{equation}
P^{\frac{1}{2}}_{R}(k)=2^{\mu-\frac{1}{2}}\frac{\Gamma(\mu)}{\Gamma(3/2)}(\mu-\frac{1}{2})^{\frac{1}{2}-\mu}\frac{1}{m^2_{pl}}\frac{H^2}{|H'|}|_{k=aH}
\end{equation}

\begin{equation}
P^{\frac{1}{2}}_g
(k)=\frac{2}{\sqrt{\pi}}2^{\nu-\frac{1}{2}}\frac{\Gamma(\nu)}{\Gamma(3/2)}(\nu-\frac{1}{2})^{\frac{1}{2}-\nu}\frac{H}{m_{pl}}|_{k=aH}
\end{equation}

The property $\mu\neq \nu$ in tachyon power-law inflation will
help us to distinguish it from that driven by ordinary scalar
field.
\subsection{The slow-roll approximation of tachyon inflation}

The slow-roll approximation has played an important role in the
inflationary cosmology. In the tachyon inflation, the situation is
much the same. The slow-roll approximation is the assumption that
the field evolution is dominated by drag from the expansion in
small parameters $\epsilon$, $|\eta|\ll 1$ about the de Sitter
limit $\epsilon = 0$.

In this approximation, it is consistent to take $\epsilon$ and
$\eta$ to be approximately constant, and the solutions are again
Hankel function of the first kind with

\begin{equation}
\mu=\frac{3}{2}+4\epsilon -2\eta
\end{equation}

\noindent and

\begin{equation}
\nu=\frac{3}{2}+\epsilon
\end{equation}

\noindent to first order in the small parameters $\epsilon$,
$\eta$. These give the final result of lowest-order

\begin{equation}\label{P12R:59}
P^{1/2}_{R}(k)=[1-(4C+2)\epsilon
+2C\eta]\frac{2}{m^2_{pl}}\frac{H^2}{|H'|}|_{k=aH}
\end{equation}

\noindent and

\begin{equation}\label{P12g:60}
P^{1/2}_g(k)=[1-(C+1)\epsilon]\frac{4}{\sqrt{\pi}}\frac{H}{m_{pl}}|_{k=aH}
\end{equation}

\noindent where $C=-2+\ln 2+\gamma \approx -0.73$ is a constant,
$\gamma$ being the Euler constant originated in the expansion of
the $\Gamma$-function.

From above discussion, we can find that the expressions of tachyon
inflation in the slow-roll approximation have obvious distinction
from the inflation driven by ordinary scalar field. It is not
surprising for the Born-Infeld action is essentially different
from the that of ordinary scalar field. Therefore, we shall obtain
different observable values for current astrophysical interest in
the tachyon case.

\section{Towards a realistic tachyon inflation}

So far, we have discussed the generic procedures for dealing with
the tachyon inflation. In this section, we will analyze a
realistic tachyon model. The form of the tachyon potential $V(T)$
may depend on the underlying (bosonic or supersymmetric) string
field theory. Recently, Kutasov, Mari$\tilde{n}$o and Moore
\cite{Kutasov} have given the tachyon potential around the maximum
in the bosonic theory,

\begin{equation}\label{V:61}
V(T)=\frac{m^4_s}{(2\pi)^3g_s}(1+\frac{T}{l_s})\exp(-\frac{T}{l_s})
\end{equation}

\noindent where $g_s$ is the string coupling, and $l_s = m_s^{-1}$
are the fundamental string length and mass scales \cite{Jones}. It
is worth noting that in the above expression (Eq.(\ref{V:61})), we
recover the dimension of the tachyon field in the units of $m_s$.
The 4D Planck mass can be rewritten by

\begin{equation}
m_{pl}^2=\frac{m_s^{d+2}r^d}{\pi g_s}
\end{equation}

\noindent where $r$ is a radius of the compactification, and $d$
is the numbers of compactified dimensions. Customarily, we can
assume $r\gg l_s$ in order to be able to use the effective 4D
field theory.However, following Sen\cite{Sen2}, the potential
should be exponential at large $T$

\begin{equation}\label{V:63}
V(T)=e^{-\frac{T}{2l_s}}
\end{equation}

\noindent For definiteness one should assume that the potential
$V(T)$ in the Born-Infeld type action is a smooth function
interpolating between two asymptotic expressions, (\ref{V:61}) at
maximum and (\ref{V:63}) at infinity.

Kofman and Linde \cite{Kofman} have shown that tachyon will
acquire immediately a matter dominated equation of state when it
rolls to its ground state at $T\rightarrow \infty$. The energy
density of matter decreases as $a^{-3}$, whereas the density of
radiation decreases as $a^{-4}$. In addition, all matter that
appears after inflation should be produced in the reheating
process. In most of reheating theory, creation of particles occurs
only when the inflation field oscillates near the minimum of its
effective potential. Since, the effective potential (\ref{V:61})
does not have any minimum at finite $T$, this mechanism does not
work. Even though a small fraction of tachyon energy is released
into radiation, it will be quickly redshifted away. Therefore if
the tachyon originally dominated the energy density of the
universe, then it never becomes radiation dominated, in
contradiction with the theory of nucleosynthesis. Summarily,
Kofman and Linde argue that above fact rules out inflationary
models where tachyon rolls to the minimum of the tachyon potential
(\ref{V:61}) at $T\rightarrow \infty$ \cite{Fairbairn}.

Alternatively, we here consider the possibility to achieve
tachyonic inflation by assuming that the inflation occurs near the
top of tachyon potential (\ref{V:61}). We assume that the
displacement of the field from its stationary value $T_0$ is small
compared to the Planck scale, $T-T_0\ll m_{pl}$, and $T_0=0$ for
tachyon potential (\ref{V:61}). We argue that slow-roll
approximation is inconsistent in this case. We may prove this
argument by contradiction. The slow-roll approximation is an
assumption that the tachyon field evolution is dominated by
dragging from expansion, $\ddot{T}\simeq 0$, so we have

\begin{equation}
\epsilon=
(2\pi)^2(\frac{r}{l_s})^d\frac{(\frac{T}{l_s})^2}{(1+\frac{T}{l_s})^3}e^{\frac{T}{l_s}}
\end{equation}

\noindent and

\begin{equation}
\eta=\pi^2(\frac{r}{l_s})^d\frac{[3(\frac{T}{l_s})^2-4]}{(1+\frac{T}{l_s})^3}e^{\frac{T}{l_s}}
\end{equation}

\noindent where the parameter $\eta$ becomes large near the top of
the tachyon potential, indicating a breakdown of the slow-roll
assumption. However, in the region

\begin{equation}
0<T\ll
(2\pi)^{\frac{3}{2}g_s^{\frac{1}{2}}}(\frac{m_{pl}}{m_s^2}),
\end{equation}

\noindent the Hamilton-Jacobi equation (\ref{HJE:8}) can be
reduced to

\begin{equation}
[1-\frac{1}{3}\epsilon(T)]^{\frac{1}{2}}=\frac{V(T)}{V(T_0)}
\end{equation}

\noindent where $T_0$ is a stationary point of tachyon field. The
parameters $\epsilon$ and $\eta$ can be written as

\begin{equation}
\epsilon(T)=\frac{3}{2}[1-(1+\frac{T}{l_s})^2\exp(-\frac{2T}{l_s})]
\end{equation}

\noindent and

\begin{equation}
\eta(T)=\epsilon-\frac{\epsilon'l_s}{4\sqrt{\epsilon}}\bigg[\frac{(12-8\epsilon)m_{pl}^4}{3V(T)}\bigg]^{\frac{1}{4}}
\end{equation}

\noindent where the second term dominates, which is equivalent to
$|\eta|\gg \epsilon$. The number of e-folds $N$ is given by

\begin{eqnarray}
N(T,T_e)&=&\int^{T_e}_{T}\bigg[\frac{4\kappa^2V^2(T)}{\epsilon^2(T)(81-54\epsilon(T))}\bigg]^{\frac{1}{4}}dT\nonumber\\
        &=&\frac{2\sqrt{3}}{g(2\pi)^{\frac{3}{2}}}\frac{m_s}{m_{pl}\sqrt{g_s}}\int^{x_e}_{x}\frac{dx}{\sqrt{1-(1+x)^2\exp(-2x)}}
\end{eqnarray}

\noindent where $x_e=\frac{T_e}{l_s}\approx 1.44$ and
$x=\frac{T}{l_s}$. In this model, the tachyon field must initially
be displaced from $T_0=0$, and we denote the initial value of $T$
by $T_i$. Following Linde's viewpoint of "chaotic inflation"
\cite{Linde}, we envision that the initial distribution of $T_i$
is "chaotic", with $T_i$ lacking on different values in different
region of the Universe. The number of e-folds must be sufficiently
large to solve the major cosmological problems. For our universe,
we usually assume $N(T,T_e)\approx 50$. Therefore, the initial
value $T_i$ should be

\begin{equation}
T_i\approx
10^{-5}m_s^{-1}\exp(-\frac{300\sqrt{6g_s}\pi^{\frac{3}{2}}m_{pl}}{m_s}).
\end{equation}

\noindent It is extremely close to the stationary point of the
tachyon field $T_0$ and depends on the parameters of string
theory, $m_s$ and $g_s$. If we take $m_s/\sqrt{g_s}=
10^{-1}m_{pl}$, $m_s/\sqrt{g_s}= 10^{-2}m_{pl}$, and
$m_s/\sqrt{g_s}= 10^{-3}m_{pl}$, the value of $\ln T_i$ is
$(\frac{1}{2}\ln g_s -40888)$, $(\frac{1}{2}\ln g_s -409157)$ and
$(\frac{1}{2}\ln g_s -4091838)$ respectively.

Next, one can introduce a new variable $y\equiv \frac{\kappa}{aH}$
to replace conformal time $\tau$ in mode equation, which
corresponds to one of conventional inflation \cite{Stewart}.
However, there are obvious difference between them. It is easy to
find

\begin{equation}
dy=-k[1-\frac{2}{3}(\frac{H'}{H^2})^2]d\tau
\end{equation}

\noindent and the mode equations reduced to

\begin{equation}\label{yne:72}
y^2(1-\epsilon)^2\frac{d^2v_k}{dy^2}+4y\epsilon(\epsilon-\eta)\frac{dv_k}{dy}+(y^2-2+\epsilon)v_k=0
\end{equation}

\noindent and

\begin{equation}\label{yme:73}
y^2(1-\epsilon)^2\frac{d^2u_k}{dy^2}+4y\epsilon(\epsilon-\eta)\frac{du_k}{dy}+[y^2-F(\epsilon,
\eta, \xi)]u_k=0
\end{equation}

\noindent where

\begin{equation}
F(\epsilon, \eta,
\xi)=2(1+4\epsilon-3\eta+9\epsilon^2-14\epsilon\eta+2\eta^2+\frac{1}{2}\xi^2).
\end{equation}

\noindent In the approximately de Sitter limit $\epsilon\ll 1$,
$|\eta|\gg \epsilon$ and

\begin{equation}
\eta=-\sqrt{\frac{3}{8}}\bigg(\frac{2\pi}{g_s}\bigg)^{\frac{1}{4}}\bigg(\frac{r}{l_s}\bigg)^{\frac{d}{2}}.
\end{equation}

\noindent Eq.(\ref{yme:73}) can be expressed as a bessel equation

\begin{equation}
[\frac{d^2}{dy^2}+1-\frac{\mu^2-\frac{1}{4}}{y^2}]u_k=0
\end{equation}

\noindent where $\mu=\frac{3}{2}-2\eta$. The spectrum of curvature
perturbation is

\begin{equation}
P^{\frac{1}{2}}_{R}(k)=2^{\mu-1/2}\frac{\Gamma(\mu)}{\Gamma(3/2)}\frac{H}{m_{pl}^2\sqrt{\epsilon}}|_{k=aH},
\end{equation}

\noindent and the scalar spectral index is

\begin{equation}
n_R-1\equiv \frac{d\ln(P_R)}{d\ln(k)}=2\eta.
\end{equation}

On the other hand, Eq.(\ref{yne:72}) can be solved in a fashion
similar to the scalar mode equation (\ref{yme:73}). However, in
the $\epsilon \ll |\eta|$ case, tensor fluctuations become
negligible relative to scalar fluctuations.

By now, the inflationary universe is generally recognized to be
the most likely scenario that explains the origin of the Big Bang.
So far, its predictions of the flatness of the universe and the
almost scale-invariant power spectrum of the density perturbation
that seeds structure formation are in good agreement with the
cosmic microwave  background (CMB) observations. The key data of
CMB are the density perturbation magnitude measured by COBE
\cite{Smoot} and its power spectrum index $n$ \cite{Lee} which
leads us to conclude that the string coupling generically should
be very strong.

\section{Summary and remarks}

In this paper, we have shown a general approach to characterizing
inflationary theory which based on the rolling tachyon dynamics.
In this approach, the standard slow-rolling approximation
corresponds to taking $\epsilon\simeq 0$ and the corresponding
solution of the Hamilton-Jacobi equation is $\epsilon
=\frac{1}{2\kappa V}\bigg(\frac{V'(T)}{V(T)}\bigg)^2$. However, in
the limit $T-T_0\ll m_{pl}/\sqrt{V(T_0)}$, a general non-slow-roll
solution exists, $\epsilon =
\frac{3}{2}\bigg[1-\frac{V^2(T)}{V^2(T_0)}\bigg]$. We calculate
the spectrum of the scalar and tensor perturbation and show an
exact solution of the spectrum of perturbation for power-law
tachyon inflation. We apply the general formalism beyond the
slow-roll approximation to the string motivated tachyon inflation,
which allows one to argue that rolling tachyon provides an
possibility of the origin of inflation.

On the other hand, the present experimental data does not
determine the values of $m_s$ and $g_s$. Therefore, a careful
analysis in realistic string models may be helpful for determining
the string parameters by using the CMB observations.

The Born-Infeld action is quite different from that of the
ordinary scalar field. Therefore, although the expressions of
tachyon inflation correspond to those of the inflation driven by
the ordinary scalar field, there is obvious difference between
them, which can not be neglected.

\vspace{5mm} \noindent {\bf Acknowledgments}

This work is partially supported by National Nature Science
Foundation of China, National Doctor Foundation of China under
Grant No. 19999025110, and Foundation of Shanghai Development for
Science and Technology 01JC1435.

\end{document}